\documentclass[12pt,a4paper]{article}
\usepackage{amsmath}
\usepackage{amsfonts}
\usepackage{graphicx}
\usepackage{cite}

\def\sf#1#2{\langle #1, #2\rangle}

\def\({\left(}
\def\){\right)}

\newcommand{\nn}{\nonumber}
\newcommand{\Eqn}[1]{&\hspace{-0.5em}#1\hspace{-0.5em}&}
\renewcommand{\[}{\begin{equation}}
\renewcommand{\]}{\end{equation}}
\newcommand{\eqb}{\begin{eqnarray}}
\newcommand{\eqe}{\end{eqnarray}}

\newcommand{\bbR}{{\mathbb R}}
\newcommand{\bbZ}{{\mathbb Z}}

\newcommand{\alg}[1]{\mathfrak{#1}}
\newcommand{\grp}[1]{\mathrm{#1}}

\newcommand{\bp}{{\bar{p}}}
\newcommand{\bw}{{\bar{w}}}
\newcommand{\bz}{{\bar{z}}}
\newcommand{\bZ}{{\bar{Z}}}
\newcommand{\bzeta}{{\bar{\zeta}}}

\newcommand{\da}{{\dot{a}}}
\newcommand{\db}{{\dot{b}}}
\newcommand{\dalpha}{{\dot{\alpha}}}
\newcommand{\dbeta}{{\dot{\beta}}}

\newcommand{\hs}{{\hat{s}}}
\newcommand{\hB}{{\hat{B}}}
\newcommand{\hS}{{\hat{S}}}
\newcommand{\hW}{{\hat{W}}}
\newcommand{\halpha}{{\hat{\alpha}}}
\newcommand{\heta}{{\hat{\eta}}}
\newcommand{\hpsi}{{\hat{\psi}}}

\newcommand{\iLambda}{\mathnormal{\Lambda}}

\newcommand{\talpha}{{\tilde{\alpha}}}

\def\cdots {\cdot\cdot\cdot}
\def\pint#1 {- \!\!\!\!\!\!\!\! \,\int_{#1}}
\def\semiket#1  { \, #1 \, \rangle \, }

\def\abs#1      {  \, \vert #1 \vert \,   }
\def\Im#1    { \, {\rm Im } \, #1  }
\def\Re#1    { \, {\rm Re}  \, #1  }

\def\trans#1  { {}^{t} \! \vec{#1} }
\def\binom#1#2 { \vecii{ {}_{#1} }{\raisebox{.5ex}{$ {}^{#2} $}} }
\def\sqbinom#1#2 { \Bigl(\begin{array}{c} {}_{#1}
                       \\ \raisebox{.5ex}{${}^{#2}$}
                         \end{array}\Bigr)^2  }

\def\r12    {\frac{r_1}{r_2}}

\def\vecii#1#2     {{ #1 \choose #2 }  }
\def\veciii#1#2#3  {\left(\begin{array}{c}#1\\#2\\#3\end{array}\right)}
\def\matrixii#1#2#3#4            {\Bigl( \begin{array}{cc}#1&#2\\#3&#4
                                     \end{array} \Bigr) }
\def\matrixiii#1#2#3#4#5#6#7#8#9 {\left(\begin{array}{ccc}#1&#2&#3\\
                                  #4&#5&#6\\#7&#8&#9\end{array}\right)}

\renewcommand{\thesection}
  {\arabic{section}.\hspace{-.5em}}
\renewcommand{\thesubsection}
  {\arabic{section}.\arabic{subsection}.\hspace{-.5em}}
\renewcommand{\thesubsubsection}
  {\arabic{section}.\arabic{subsection}.\arabic{subsubsection}.\hspace
                                                               {-.5em}}

\makeatletter
\renewcommand\section{
  \@startsection{section}{3}{\z@}%
  {-3.25ex\@plus -1ex \@minus -.2ex}%
  {1.5ex \@plus .2ex}%
  {\normalfont\normalsize\bfseries\mathversion{bold}}}
\renewcommand\subsection{
  \@startsection{subsection}{3}{\z@}%
  {-3.25ex\@plus -1ex \@minus -.2ex}%
  {1.5ex \@plus .2ex}%
  {\normalfont\normalsize\bfseries\mathversion{bold}}}
\renewcommand\subsubsection{
  \@startsection{subsubsection}{3}{\z@}%
  {-3.25ex\@plus -1ex \@minus -.2ex}%
  {1.5ex \@plus .2ex}%
  {\normalfont\normalsize\itshape}}
\makeatother

\makeatletter \@addtoreset{equation}{section} \makeatother
\renewcommand{\theequation}{\arabic{section}.\arabic{equation}}

\makeatletter
\renewcommand{\appendix}{
\renewcommand{\thesection}{Appendix \Alph{section}.\hspace{-.5em}}
\renewcommand{\thesubsection}
  {\Alph{section}.\arabic{subsection}.\hspace{-.5em}}
\renewcommand{\thesubsubsection}
  {\Alph{section}.\arabic{subsection}.\arabic{subsubsection}.\hspace
                                                             {-.5em}}
\@addtoreset{equation}{subsection}
\renewcommand{\theequation}{\Alph{section}.\arabic{equation}}
\setcounter{section}{0}}
\makeatother

\begin{document}
%
\def\papertitlepage{\baselineskip 3.5ex \thispagestyle{empty}}
\def\preprinumber#1#2#3#4{\hfill
\begin{minipage}{1.2in}
#1 \par\noindent #2 \par\noindent #3 \par\noindent #4
\end{minipage}}
\renewcommand{\thefootnote}{\fnsymbol{footnote}}
\newcounter{aff}
\renewcommand{\theaff}{\fnsymbol{aff}}
\newcommand{\affiliation}[1]{
  \setcounter{aff}{#1} $\rule{0em}{1.2ex}^\theaff\hspace{-.4em}$}
%
%
\papertitlepage
\setcounter{page}{0}
\preprinumber{RIKEN-TH-185}{TIT/HEP-602}{UTHEP-604}{arXiv:1002.2941}
\vskip 2ex
~
\vfill
\begin{center}
{\large\bf\mathversion{bold}
Thermodynamic Bethe Ansatz Equations\\
for Minimal Surfaces in $AdS_3$}
\end{center}
\vfill
\baselineskip=3.5ex
\begin{center}
  Yasuyuki Hatsuda\footnote[1]{\tt hatsuda@riken.jp}, 
  Katsushi Ito\footnote[2]{\tt ito@th.phys.titech.ac.jp},
  Kazuhiro Sakai\footnote[3]{\tt sakai@phys-h.keio.ac.jp} and
  Yuji Satoh\footnote[4]{\tt ysatoh@het.ph.tsukuba.ac.jp}\\

\vskip 2ex
  \affiliation{1}
  {\it Theoretical Physics Laboratory, RIKEN}\\
  {\it Saitama 351-0198, Japan}\\
 
\vskip 1ex
  \affiliation{2}
  {\it Department of Physics, Tokyo Institute of Technology}\\
  {\it Tokyo 152-8551, Japan}\\

\vskip 1ex
  \affiliation{3}
  {\it Research and Education Center for Natural Sciences}\\
  {\it and Hiyoshi Department of Physics, Keio University}\\
  {\it Yokohama 223-8521, Japan}\\

\vskip 1ex
  \affiliation{4}
  {\it Institute of Physics, University of Tsukuba}\\
  {\it Ibaraki 305-8571, Japan}

\end{center}
\vfill
%
\baselineskip=3.5ex
\begin{center} {\bf Abstract} \end{center}

We study classical open string solutions
with a null polygonal boundary in $AdS_3$
in relation to gluon scattering amplitudes
in ${\cal N}\!=\!4$ super Yang--Mills at strong coupling.
We derive in full detail the set of integral equations
governing the decagonal and the dodecagonal
solutions and identify them
with the thermodynamic Bethe ansatz equations
of the homogeneous sine-Gordon models.
By evaluating the free energy in the conformal limit
we compute the central charges,
from which we observe general correspondence between
the polygonal solutions in $AdS_n$ and generalized parafermions.

\vskip 2ex
\vspace*{\fill}
\noindent
February 2010
\setcounter{page}{0}
\newpage
\renewcommand{\thefootnote}{\arabic{footnote}}
\setcounter{footnote}{0}
\setcounter{section}{0}
\baselineskip = 3.5ex
\pagestyle{plain}
%

\section{Introduction}

Recently there has been much interest in computing
 gluon scattering amplitudes in ${\cal N}\!=\!4$ super Yang--Mills
theory at strong coupling by using AdS/CFT correspondence.
The amplitude is dual to the Wilson loop with light-like segments
\cite{Dr}, which 
corresponds to the area of  minimal surfaces in AdS with the same
 boundary \cite{Alday:2007hr}.

In \cite{Alday:2007hr}, the minimal surface for the 4-point amplitude
has been obtained by solving the Euler-Lagrange equation in the static
gauge. 
The minimal surfaces in AdS are further studied in
\cite{various,Sakai:2010eh}
but it is a very difficult problem to extend the 4-point
solution to the 
general $n$-point amplitudes.
It is important to evaluate the area for the $n$-point
amplitudes in order to determine the remainder function, which
represents
deviation from the conjectured BDS formula of the multi-loop amplitudes
\cite{BDS}.
The remainder functions, which are  functions of cross-ratios of
momenta, are shown to exist in the 6-point amplitudes at two-loop level
\cite{Dr2} and 
evaluated numerically\cite{nume} and explicitly\cite{Smir}.

Recently, there has been remarkable progress in obtaining exact
solutions of minimal surfaces with a null polygonal boundary in AdS.
It is shown that 
the equations for the minimal surface in $AdS_3$ can be reduced to
the $\grp{SU}(2)$ Hitchin equations \cite{Alday:2009yn}.
The minimal area is obtained by finding the Stokes data of the
associated linear problem, which is studied in detail by
Gaiotto, Moore and Neitzke \cite{Gaiotto:2008cd,Gaiotto:2009hg}.
The explicit formula for the area of the minimal surface for
the 8 sided polygon has been obtained \cite{Alday:2009yn}.
This is  further generalized to the $AdS_4$
\cite{Alday:2009dv}, \cite{Burrington:2009bh}
and the $AdS_5$ case \cite{Alday:2009dv}. In the $AdS_5$ case,
the minimal area problem is shown to be equivalent to solving the
$\grp{SU}(4)$ Hitchin system.
Motivated by the connection between the solution of the associated 
linear problem and the Thermodynamic Bethe Ansatz (TBA) integral
equations \cite{Gaiotto:2008cd}, Alday, Gaiotto and Maldacena found that
the minimal area of the 6 sided polygon is evaluated by the free energy
of the TBA equations of the $A_3$ integrable theory \cite{Alday:2009dv}.
TBA equations also appear in the study of the spectral problem
in AdS/CFT correspondence \cite{TBAinAdSCFT}.

The TBA equations \cite{Zamolodchikov:1989cf}
have been studied extensively in the investigation of
the massive integrable field theory and its relevant perturbed conformal
field theory (CFT). It will be a quite interesting problem to 
study the role of the TBA equations in the minimal area problem in AdS. 

In this paper we study minimal surfaces with a null polygonal boundary 
in AdS. 
We focus on the minimal surfaces with 
a $2n$ sided polygonal boundary in $AdS_3$.
We determine the integral equations explicitly in the case of the
decagon and the dodecagon.
We find that the integral  equations fit precisely in the general form 
proposed by Gaiotto, Moore and Neitzke \cite{Gaiotto:2008cd}.
We identify  the present integral equations with the TBA equations
of the homogeneous sine-Gordon model
\cite{FernandezPousa:1996hi,FernandezPousa:1997zb}.
The free energy of the TBA system is related to the regularized area
\cite{Alday:2009dv}.
We then evaluate the free energy and compute the central charges
for the decagon and dodecagon.
We find that the regularized areas precisely match those obtained from
the central charges in
the CFT limit of the TBA system.
We generalize these results to the general $2n$ sided polygons in
$AdS_3$ and argue that relevant CFTs are identified with 
generalized parafermion theories
for $\grp{SU}(n-2)_{2}/\grp{U}(1)^{n-3}$.
We comment on the case of $AdS_5$.

This paper is organized as follows.
In sect.~2, we review the construction of open string solution in
$AdS_3$ \cite{Alday:2009yn} and discuss the Stokes data of
the associated linear problem.
In sect.~3, we analyze the Riemann--Hilbert problem and introduce the 
functional variables with simple asymptotics for the decagon and
the dodecagon cases. 
We derive the integral equations and identify them with
the TBA equations of the homogeneous sine-Gordon models.
In sect.~4, we study the free energy of the homogeneous sine-Gordon
models
and their CFT limit. 
We compare the central charges and the regularized areas.
In sect. 5, we present conclusions and discussion.

\section{Minimal surfaces with a null polygonal boundary in $AdS_3$}

\subsection{The linear problem}

In this paper we consider classical open string solutions in $AdS_3$
with a Euclidean world-sheet.\footnote{
In this subsection we basically follow the notation
of \cite{Alday:2009yn}.}
Let $z$ be a complex coordinate parametrizing the world-sheet.
Let $\vec{Y}=(Y_{-1},Y_0,Y_1,Y_2)^{\rm T}\in\bbR^{2,2}$
denote the global coordinate parametrizing the 
$AdS_3$ spacetime. The $AdS_3$ is given as a hypersurface
\eqb\label{AdSeq}
\vec{Y}\cdot\vec{Y}:=-Y_{-1}^2-Y_0^2+Y_1^2+Y_2^2=-1
\eqe
in $\bbR^{2,2}$.
Inner product of two vectors $\vec{A},\vec{B}\in\bbR^{2,2}$
is defined as
\eqb
\vec{A}\cdot\vec{B}=\eta_{ij}A^iB^j,\qquad
\eta_{ij}
={\rm diag}(-1,-1,+1,+1).
\eqe
The solution is given by
a function $\vec{Y}(z,\bz)$
satisfying the constraint (\ref{AdSeq}),
the classical equations of motion
\[\label{AdSEOM}
\vec{Y}_{z\bz}-(\vec{Y}_z\cdot\vec{Y}_\bz)\vec{Y}=0
\]
and the Virasoro constraints
\[\label{Virasoro}
\vec{Y}_z^2=\vec{Y}_\bz^2 =0.
\]
Here we abbreviate the world-sheet derivatives as
$\partial_z\vec{Y}=\vec{Y}_z,
\ \partial_\bz\vec{Y}=\vec{Y}_\bz$.

Let us introduce the following notations
\eqb
e^{2\alpha}\Eqn{:=}\frac{1}{2}\vec{Y}_z\cdot\vec{Y}_\bz\,,\\
N_i \Eqn{:=} \frac{1}{2}e^{-2\alpha}\epsilon_{ijkl}Y^j Y^k_z Y^l_\bz\,,
\qquad \epsilon_{(-1)012}=+1,\\
p\Eqn{:=}-\frac{1}{2}\vec{N}\cdot\vec{Y}_{zz}\,.
\eqe
The pseudo vector $\vec{N}$ is chosen so that
\[
\vec{N}\cdot\vec{Y}=\vec{N}\cdot\vec{Y}_z=\vec{N}\cdot\vec{Y}_\bz=0,
\qquad
\vec{N}\cdot\vec{N}=1.
\]
Using (\ref{AdSeq}), (\ref{AdSEOM}), (\ref{Virasoro}), one can show that
\eqb
\overline{\alpha}\Eqn{=}\alpha,\qquad
\overline{\vec{N}}=-\vec{N},\\
p_\bz\Eqn{=}0,
\eqe
{\it i.e.}, $\alpha(z,\bz)$ is real-valued and $p(z)$ is holomorphic
for the string solutions.

One can consider a moving frame basis
spanned by the following vectors
\[
\vec{q}_1=\vec{Y},\qquad 
\vec{q}_2=e^{-\alpha}\vec{Y}_\bz,\qquad 
\vec{q}_3=e^{-\alpha}\vec{Y}_z,\qquad 
\vec{q}_4=\vec{N}.
\]
We recast them into the following form
\[\label{Wcomponents}
W_{\alpha\dalpha,a\da}
=\left(\begin{array}{cc}
W_{11,a\da}&W_{12,a\da}\\
W_{21,a\da}&W_{22,a\da}
\end{array}
\right)
=\frac{1}{2}
\left(\begin{array}{cc}
(q_1+q_4)_{a\da}&(q_2)_{a\da}\\
(q_3)_{a\da}&(q_1-q_4)_{a\da}
\end{array}
\right),
\]
where $\alpha,\dalpha=1,2$ denote internal spinor indices
while $a,\da=1,2$ denote spacetime spinor indices.
In the r.h.s.~of (\ref{Wcomponents})
the 4-vectors $\vec{q}_j$ are expressed in the spinor notation.
In this notation $\vec{Y}$, for example, is expressed as
\[\label{Ycomponents}
Y_{a\da}
=\left(\begin{array}{cc}
Y_{11}&Y_{12}\\
Y_{21}&Y_{22}
\end{array}\right)
=\left(\begin{array}{cc}
Y_{-1}+Y_2&Y_1-Y_0\\
Y_1+Y_0&Y_{-1}-Y_2
\end{array}\right).
\]
Evolution of $W_{\alpha\dalpha,a\da}$ is described by
the following set of linear equations
\eqb
\label{Wlineq1}
\partial_z W_{\alpha\dalpha,a\da}
+(B_z^{\rm L})_\alpha{}^\beta W_{\beta\dalpha,a\da}
+(B_z^{\rm R})_\dalpha{}^\dbeta W_{\alpha\dbeta,a\da}\Eqn{=}0,\\
\label{Wlineq2}
\partial_\bz W_{\alpha\dalpha,a\da}
+(B_\bz^{\rm L})_\alpha{}^\beta W_{\beta\dalpha,a\da}
+(B_\bz^{\rm R})_\dalpha{}^\dbeta W_{\alpha\dbeta,a\da}\Eqn{=}0,
\eqe
where
\eqb\label{BzLR}
B_z^{\rm L}\Eqn{=}B_z(1),\qquad B_z^{\rm R}=UB_z(i)U^{-1},\\
B_\bz^{\rm L}\Eqn{=}B_\bz(1),\qquad B_\bz^{\rm R}=UB_\bz(i)U^{-1}
\eqe
with
\eqb\label{Bz}
B_z(\zeta)\Eqn{=}\phantom{-}\frac{\alpha_z}{2}
\left(\begin{array}{cc}
1&0\\
0&-1
\end{array}\right)
-\frac{1}{\zeta}
\left(\begin{array}{cc}
0&e^\alpha\\
e^{-\alpha}p&0
\end{array}\right),
\\
\label{Bbz}
B_\bz(\zeta)\Eqn{=}-\frac{\alpha_\bz}{2}
\left(\begin{array}{cc}
1&0\\
0&-1
\end{array}\right)
-\zeta
\left(\begin{array}{cc}
0&e^{-\alpha}\bp\\
e^\alpha&0
\end{array}\right),
\\
\label{Uform}
U\Eqn{=}\left(\begin{array}{cc}
0&e^{\frac{\pi i}{4}}\\
e^{\frac{3\pi i}{4}}&0
\end{array}
\right).
\eqe
The equations of motion (\ref{AdSEOM})
as well as the constraints (\ref{AdSeq}), (\ref{Virasoro})
have been used
in deriving the particular form of $B_z,B_\bz$.
In other words, the linear equations (\ref{Wlineq1}), (\ref{Wlineq2})
with the connections given by (\ref{BzLR})--(\ref{Uform})
encode the conditions for $\vec{Y}$.

The above evolution equations exhibit a peculiar structure that
the connection decomposes into
the left and the right parts.
Moreover, since each entry of the matrix (\ref{Wcomponents})
is a null vector,
$W_{\alpha\dalpha,a\da}$ can be expressed
as a product of two spinors
\[\label{Wdecomposition}
W_{\alpha\dalpha,a\da}
=\psi_{\alpha,a}^{\rm L}\psi_{\dalpha,\da}^{\rm R}\,.
\]
Therefore,
once we fix such decomposition at one point on the $z$-plane,
$\psi_{\alpha,a}^{\rm L}$ and $\psi_{\dalpha,\da}^{\rm R}$
evolve separately over the whole $z$-plane, obeying
\eqb
\label{psiLlineq}
\partial_z \psi_{\alpha,a}^{\rm L}
+(B_z^{\rm L})_\alpha{}^\beta \psi_{\beta,a}^{\rm L}\Eqn{=}0,\qquad
\partial_\bz \psi_{\alpha,a}^{\rm L}
+(B_\bz^{\rm L})_\alpha{}^\beta \psi_{\beta,a}^{\rm L}=0,\\
\label{psiRlineq}
\partial_z \psi_{\dalpha,\da}^{\rm R}
+(B_z^{\rm R})_\dalpha{}^\dbeta \psi_{\dbeta,\da}^{\rm R}\Eqn{=}0,\qquad
\partial_\bz \psi_{\dalpha,\da}^{\rm R}
+(B_\bz^{\rm R})_\dalpha{}^\dbeta \psi_{\dbeta,\da}^{\rm R}=0.
\eqe
The original string solutions
are constructed from solutions of these equations.
Note that the solutions have to satisfy not only
the evolution equations (\ref{Wlineq1}), (\ref{Wlineq2}),
but also the normalization condition
\eqb
\epsilon^{ab}\epsilon^{\da\db}
W_{\alpha\dalpha,a\da}W_{\beta\dbeta,b\db}
=\epsilon_{\alpha\beta}\epsilon_{\dalpha\dbeta}
\eqe
and the reality condition
\eqb
\overline{W_{\alpha\dalpha,a\da}}=W_{\alpha\dalpha,a\da}.
\eqe
Such solutions are obtained if one slightly generalize
(\ref{Wdecomposition}) as
\[
W_{\alpha\dalpha,a\da}
=M_{a\da,b\db}
\psi_{\alpha,b}^{\rm L}\psi_{\dalpha,\db}^{\rm R}\,
\]
and determine the constant $M_{a\da,b\db}$ accordingly.\footnote{
The discussion is similar to that in the case of finite-gap solutions
\cite{Sakai:2010eh}.}
For later purpose we adopt the normalization conditions
\eqb
\psi_{a}^{\rm L}\wedge\psi_{b}^{\rm L}\equiv
\epsilon^{\beta\alpha}\psi_{\alpha,a}^{\rm L}\psi_{\beta,b}^{\rm L}
=\epsilon_{ab}\,,\qquad
\psi_{\da}^{\rm R}\wedge\psi_{\db}^{\rm R}\equiv
\epsilon^{\dbeta\dalpha}
  \psi_{\dalpha,\da}^{\rm R}\psi_{\dbeta,\db}^{\rm R}
=\epsilon_{\da\db}\,,
\eqe
where $\psi_{a}^{\rm L}$ for $a=1,2$ and
$\psi_{\da}^{\rm R}$ for $\da=1,2$
are two linearly independent solutions of
(\ref{psiLlineq}) and (\ref{psiRlineq}), respectively.

The linear problems
(\ref{psiLlineq}) and (\ref{psiRlineq})
can be promoted to a family of linear problems with
the general spectral parameter $\zeta$
\[\label{lineq_z}
\Bigl(\partial_z+B_z(\zeta)\Bigr)\psi(z,\bz;\zeta)=0,\qquad
\Bigl(\partial_\bz+B_\bz(\zeta)\Bigr)\psi(z,\bz;\zeta)=0,
\]
where the connections are given by (\ref{Bz}), (\ref{Bbz}).
From now on we use matrix notations for indices $\alpha,\dalpha$,
where $B_z,B_\bz$ denote $2\times 2$ matrices
as well as $\psi$ a $2$-component column vector.

By using the variable such that
\[
dw=\sqrt{p(z)}dz
\]
with the redefinition
\[
\halpha=\alpha-\frac{1}{4}\log p\bp
\]
and the gauge transformation
\[
\hpsi=g\psi,
\]
\[
g=e^{i\frac{\pi}{4}\sigma^3}e^{i\frac{\pi}{4}\sigma^2}
e^{\frac{1}{8}\log\frac{p}{\bp}\sigma^3}=
\left(\begin{array}{cc}
\frac{1+i}{2}&\frac{1+i}{2}\\
\frac{-1+i}{2}&\frac{1-i}{2}
\end{array}\right)
\left(\begin{array}{cc}
\left(\frac{p}{\bp}\right)^{\frac{1}{8}}&0\\
0&\left(\frac{p}{\bp}\right)^{-\frac{1}{8}}
\end{array}\right),
\]
one can completely remove $p(z)$ from
the equations (at the price of having complicated branch cut structure).
One then obtains
\[\label{lineq_w}
\left(\partial_w+\hB_w\right)\hpsi=0,\qquad
\left(\partial_\bw+\hB_\bw\right)\hpsi=0,
\]
where
\eqb
\hB_w(\zeta)\Eqn{=}\phantom{-}\frac{\halpha_w}{2}
\left(\begin{array}{cc}
0&-i\\
i&0
\end{array}\right)
-\frac{1}{\zeta}
\left(\begin{array}{cc}
\cosh\halpha&i\sinh\halpha\\
i\sinh\halpha&-\cosh\halpha
\end{array}\right),
\\
\hB_\bw(\zeta)\Eqn{=}-\frac{\halpha_\bw}{2}
\left(\begin{array}{cc}
0&-i\\
i&0
\end{array}\right)
-\zeta
\left(\begin{array}{cc}
\cosh\halpha&-i\sinh\halpha\\
-i\sinh\halpha&-\cosh\halpha
\end{array}\right).
\eqe
Note that the linear differential operators
are transformed under the gauge transformation as
\eqb
\left[\partial_w+\hB_w\right]=
p^{-\frac{1}{2}}g\left[\partial_z+B_z\right]g^{-1},\quad
\left[\partial_\bw+\hB_\bw\right]=
\bp^{-\frac{1}{2}}g\left[\partial_\bz+B_\bz\right]g^{-1}.
\eqe
%

\subsection{Bases of solutions}

We are interested in open string solutions
whose boundary consists of light-like segments
forming $2n$-gons on the AdS boundary \cite{Alday:2009yn}.
We consider solutions such that $p(z)$ is a polynomial of degree
$n-2$ and $\halpha$ behaves as
\[
\halpha\to0\qquad\mbox{for}\qquad |w|\to\infty.
\]
For solutions with such $\halpha$,
equations (\ref{lineq_w}) are approximated at large $|w|$ as
\[
(\partial_w-\zeta^{-1}\sigma_3)\hpsi=0,\qquad
(\partial_\bw-\zeta\sigma_3)\hpsi=0,
\]
which have two independent solutions
\[
\heta_+ =
\left(\begin{array}{c}
e^{\left(\frac{w}{\zeta}+\bw\zeta\right)}\\ 0
\end{array}\right),\qquad
\heta_- =
\left(\begin{array}{c}
0\\ e^{-\left(\frac{w}{\zeta}+\bw\zeta\right)}
\end{array}\right).
\]
These describe the asymptotic behavior of the big
and the small solutions of (\ref{lineq_w})
in each Stokes sector.
In particular, the small solutions are uniquely specified
by the asymptotic behavior.
Let us introduce the notation $\hs_{2k-1}(w,\bw;\zeta)$
for the small solutions in each Stokes sector.
They are characterized by the following asymptotic behavior
at large $|w|$
\eqb\label{hs_asymp}
\hs_{2k-1}\simeq(-1)^{k-1}\heta_- \quad\mbox{for}\quad
w\in\hW_{2k-1},\qquad
\hs_{2k}\simeq(-1)^k\heta_+ \quad\mbox{for}\quad
w\in\hW_{2k},\qquad
\eqe
where $\hW_j$'s denote the Stokes sectors
\eqb
\hW_j:\quad
(j-\tfrac{3}{2})\pi+\arg\zeta < \arg w <(j-\tfrac{1}{2})\pi+\arg\zeta.
\eqe
We have determined the normalizations of $\hs_j$'s so that
\eqb\label{detcond}
\hs_j\wedge\hs_{j+1}\Eqn{\equiv}\det (\hs_j\ \hs_{j+1})\nn\\
\Eqn{=}1.
\eqe
(\ref{hs_asymp})--(\ref{detcond})
uniquely determine the small solutions $\hs_j$.

One can take $\hs_{j-1}$ and $\hs_j$ as the basis
of the solutions.
Then $\hs_{j+1}$ is expressed as
\[\label{betarel}
\hs_{j+1}=-\hs_{j-1}+b_j\hs_j,
\]
where the coefficient of $\hs_{j-1}$ is determined by (\ref{detcond})
and $b_j(\zeta)$ is a coefficient independent of $w,\bw$.
It can be expressed as
\[
b_j(\zeta)=\hs_{j-1}\wedge \hs_{j+1}.
\]

Next let us consider the periodicity constraint.
First let us introduce pull-back of the basis solutions
$s_j(z,\bz;\zeta)$ by
\[
s_j=g^{-1}\hs_j.
\]
Let us also introduce the notation
\[
W_j:\quad
\frac{(2j-3)\pi}{n}+\frac{2}{n}\arg\zeta 
< \arg z < \frac{(2j-1)\pi}{n}+\frac{2}{n}\arg\zeta
\]
for the Stokes sectors on the $z$-plane.
$s_j(z,\bz;\zeta)$ are solutions to the original differential
equations (\ref{lineq_z}).
Since differential operators in (\ref{lineq_z}) have no singularities
at $|z|<\infty$, solutions $\psi(z,\bz;\zeta)$ to (\ref{lineq_z})
are also regular over the whole $z$-plane. This means that
the Stokes sectors $W_{j+n}$ and $W_j$ are identified
for $2n$-gon solutions.
Therefore $s_{j+n}$ and $s_j$, which are both the small solution
in this sector, coincide up to a normalization factor
\[\label{murel}
s_{j+n}=\mu_js_j.
\]

This factor is identified with the formal monodromy as follows.
$\hs_j$ becomes a large solution in the Stokes sector
$\hW_{j-1}$ and $\hW_{j+1}$.
In the Stokes sector $\hW_j$, both $s_{j-1}$ and $s_{j+1}$
grow with the same largest exponent.
In this way we see that $s_j$ with even $j$
grow with the same largest exponent
while $s_j$ with odd $j$ grow
with the other exponent.
Therefore by going around the $z$-plane twice,
one can evaluate $\mu_j$ as
\[
\mu_j^2=\exp(S_j(e^{4\pi i}z)-S_j(z)),
\]
where $S_j$ denotes the corresponding largest exponent.
The explicit form of the exponents
can be read from
the components of $s_j$
\eqb
\left(\begin{array}{c}
s_{j,1}\\
s_{j,2}
\end{array}\right)
\Eqn{=}
\left(\begin{array}{cc}
\left(\frac{p}{\bp}\right)^{-\frac{1}{8}}&0\\
0&\left(\frac{p}{\bp}\right)^{\frac{1}{8}}
\end{array}\right)
\left(\begin{array}{cc}
\frac{1-i}{2}&\frac{-1-i}{2}\\
\frac{1-i}{2}&\frac{1+i}{2}
\end{array}\right)
\left(\begin{array}{c}
\hs_{j,1}\\
\hs_{j,2}
\end{array}\right)\nn\\
\Eqn{=}
\left(\begin{array}{c}
\left(\frac{p}{\bp}\right)^{-\frac{1}{8}}
\left(\frac{1-i}{2}\hs_{j,1}-\frac{1+i}{2}\hs_{j,2}\right)\\[1ex]
\left(\frac{p}{\bp}\right)^{+\frac{1}{8}}
\left(\frac{1-i}{2}\hs_{j,1}+\frac{1+i}{2}\hs_{j,2}\right)
\end{array}\right).
\eqe
As $p(z)$ is a polynomial of degree $n-2$, the factor
$\left(p/\bp\right)^{1/8}$ 
contributes to the monodromy by
\[\label{phasefactor}
\left(\frac{p}{\bp}\right)^{\frac{1}{8}}\to
e^{(n-2)\pi i}\left(\frac{p}{\bp}\right)^{\frac{1}{8}}
\quad\mbox{as}\quad z\to e^{4\pi i}z
\]
at large $|z|$.
For $n$ even, the phase factor in the above equation
is trivial, but the formal monodromy receives
a contribution from the residue appearing in the $1/z$
expansion of $\ln\left(p/\bp\right)^{1/8}$.
For $n$ odd, the phase factor in (\ref{phasefactor})
gives $-1$. In this case no contribution comes from
the series expansion. Hence one obtains
\[\label{muodd}
\mu_j=\pm i\quad\mbox{for}\quad n:\mbox{odd}.
\]

In order to figure out the relation among
$b_j(\zeta)$ and $\mu$,
it is convenient to introduce the notations
\[
\hS_j=\left(\hs_j\ \hs_{j+1}\right),\quad
B_j=\left(\begin{array}{cc}
0&-1\\
1&b_j
\end{array}\right),\quad
M_j=\left(\begin{array}{cc}
\mu_j&0\\
0&\mu_{j+1}
\end{array}\right).
\]
Then (\ref{betarel}) and (\ref{murel}) are expressed as
\eqb
\hS_{j+1}\Eqn{=}\hS_jB_{j+1},\\
\hS_{j+n}\Eqn{=}\hS_jM_j.
\eqe
As $\hS_j$ is invertible, it follows that
\eqb\label{BMrel}
B_{j+1}B_{j+2}\cdots B_{j+n}=M_j.
\eqe
Since $\det B_j=1$, we see that $\det M_j=1$, namely
\eqb
\mu_{j+1}=\mu_j^{-1}.
\eqe
Therefore one can set
\eqb
\mu_{2k-1}=\mu_{2k}^{-1}=\mu\qquad k=1,2,3,\ldots.
\eqe
It also follows from (\ref{BMrel}) that
\eqb
B_{j+1}M_{j+1}=M_jB_{j+n+1},
\eqe
which gives
\eqb\label{betacycle}
b_{j+n}=\mu_{j}^{-2}b_j.
\eqe
%

\subsection{Constraints from involutions}

It can be easily checked that the pair of Dirac operators
satisfies a holomorphic involution\footnote{
This $\bbZ_2$ symmetry
corresponds to the $\bbZ_4$ symmetry
appearing in the case of $AdS_5$.
}
\eqb\label{Z2symmetry}
\sigma_2[\partial_w+\hB_w(\zeta)]\sigma_2
\Eqn{=}[\partial_w+\hB_w(-\zeta)],\quad
\sigma_2[\partial_\bw+\hB_\bw(\zeta)]\sigma_2
=[\partial_\bw+\hB_\bw(-\zeta)]\quad
\eqe
and an antiholomorphic involution
\eqb
\overline{\partial_w+\hB_w(\zeta)}
\Eqn{=}\partial_\bw+\hB_\bw(\bzeta^{-1}),\quad
\overline{\partial_\bw+\hB_\bw(\zeta)}
=\partial_w+\hB_w(\bzeta^{-1}).\quad
\eqe
The implication of these involutions is the following:
If $\hpsi(\zeta)$ is a solution
to the equations (\ref{lineq_w}),
so are $\sigma_2\hpsi(-\zeta)$ and
$\overline{\hpsi(\bzeta^{-1})}$.

Let us first examine the constraints arising
from the holomorphic involution.
As $\hs_1(w,\bw;\zeta)$ is a solution to (\ref{lineq_w}),
so is $\sigma_2\hs_1(w,\bw;e^{\pi i}\zeta)$.
It exhibits the asymptotic behavior as
\eqb
\sigma_2\hs_1(w,\bw;e^{\pi i}\zeta)
\Eqn{\simeq}-i\heta_+(w,\bw;\zeta)
\quad\mbox{for}\quad w\in \hW_2.
\eqe
Note that the asymptotic behavior appears in the Stokes sector
$\hW_2$ for $\hs_1$ with the spectral parameter $e^{\pi i}\zeta$.
As the above solution is the small solution in $\hW_2$,
it should be identified with $\hs_2$ as
\eqb
\sigma_2\hs_1(w,\bw;e^{\pi i}\zeta)=i\hs_2(w,\bw;\zeta).
\eqe
Similarly, one can show that
\eqb
\sigma_2\hs_j(w,\bw;e^{\pi i}\zeta)=i\hs_{j+1}(w,\bw;\zeta).
\eqe
It then follows that
\[\label{tbetaHcond}
b_j(e^{\pi i}\zeta)=b_{j+1}(\zeta).
\]

The antiholomorphic involution implies that
for each $j$, $\overline{\hs_j(w,\bw;\bzeta^{-1})}$ is a solution.
Analysis of the asymptotic behavior tells us that
it is the small solution in the Stokes sector $\hW_j$.
Thus one can identify it as
\[
\overline{\hs_j(w,\bw;\bzeta^{-1})}=\hs_j(w,\bw;\zeta).
\]
It then follows that
\[\label{tbetaAHcond}
\overline{b_j(\bzeta^{-1})}=b_j(\zeta).
\]
The constraints (\ref{tbetaHcond}) and (\ref{tbetaAHcond})
are peculiar to the current Hitchin system which
originates from the classical strings in $AdS_3$.

\section{Thermodynamic Bethe ansatz equations}

In this section we derive integral equations
which characterize the minimal surfaces
with a null polygonal boundary in $AdS_3$.
As demonstrated in \cite{Alday:2009dv} in the case of
hexagonal solutions in $AdS_5$,
the functional form of $b_j(\zeta)$ is fully determined
by a Riemann--Hilbert problem.
To construct the Riemann--Hilbert problem,
one needs boundary conditions
in addition to the relations and the constraints for $b_j$ derived
in the last section.
The boundary conditions are given by
asymptotic behavior for $|\zeta|\to\infty$ and for $|\zeta|\to 0$. 
What makes the story nontrivial is that
each $b_j(\zeta)$ exhibits simple
asymptotics only in some particular
angular sectors in the $\zeta$ plane.
In order to write down the Riemann--Hilbert problem
in a simple form, one therefore introduces
new functional variables $\chi_j(\zeta)$,
which have a simple asymptotics in all the angular sectors but
have discontinuities along some of the semi-infinite border lines.
A reasonable definition of $\chi_j(\zeta)$ was
presented in \cite{Gaiotto:2009hg}
as the Fock--Goncharov coordinates \cite{Fock:2006}.
The coordinates are defined for every WKB triangulation,
which is uniquely determined given the value of $\zeta$.
One can then figure out
the explicit relations between $\chi_j$ and $b_j$.
The behavior of $\chi_j$ at discontinuities
is described by the periodicity condition of $b_j$
discussed in the last section.
Below we illustrate using some simple examples
how the Riemann--Hilbert problem
is constructed from the data of $p(z)$ and
the constraints of $b_j$.

The Riemann--Hilbert problem for $\chi_j$ is written
in the form of integral equations.
It was pointed out that
the integral equations possess the structure
of TBA equations \cite{Gaiotto:2008cd}.\footnote{A connection
between TBA systems and certain ordinary differential
equations has been found \cite{DoreyTateo}, where cross-ratios turn
out to play an interesting role\cite{Gliozzi:1995wq, DoreyTateo}.
For generalizations, see for example\cite{TBAODE}.}
However, it was not clear what models are described
by these equations in practice.
We find that the TBA equations in the present cases
are identified with those of the homogeneous sine-Gordon model.

\subsection{Decagon solutions ($n=5$)}

\subsubsection{Periodicity condition}

Let us now focus on the case of decagon ($n=5$).
The condition (\ref{BMrel}) for $j=0$
is written down for each component as
\eqb
\mu^{-1}\Eqn{=}
b_2+b_4-b_2b_3b_4,\\
0\Eqn{=}
-1+b_2b_3+b_2b_5+b_4b_5
-b_2b_3b_4b_5,\\
0\Eqn{=}
1-b_1b_2-b_1b_4-b_3b_4
+b_1b_2b_3b_4,\\
\mu\Eqn{=}
b_1+b_3+b_5
-b_1b_2b_3-b_1b_2b_5
-b_1b_4b_5-b_3b_4b_5
+b_1b_2b_3b_4b_5.
\eqe
These can be simplified as
\eqb
b_1b_2\Eqn{=}1-\mu b_4,\\
b_2b_3\Eqn{=}1-\mu^{-1}b_5,\\
b_3b_4\Eqn{=}1-\mu^{-1}b_1,\\
b_4b_5\Eqn{=}1-\mu b_2.
\eqe
Similar relations are obtained from (\ref{BMrel}) with other $j$'s.
By introducing a new notation by
\eqb
\beta_{2k-1}=\mu^{-1}b_{2k-1},\quad \beta_{2k}=\mu b_{2k},
\eqe
and using (\ref{muodd}), these relations can be concisely written
as
\eqb\label{connectivity}
\beta_{j}\beta_{j+1}=1-\beta_{j+3}.
\eqe
Note that in terms of $\beta_j$'s,
with the help of (\ref{muodd}),
the relation (\ref{betacycle}) is simplified as
\eqb
\beta_{j+5}=\beta_{j}.
\eqe

We are now in a position to consider the integral equations.
The form of the integral equations is
characterized by the polynomial $p(z)$ and the connectivity
condition (\ref{connectivity}).
In the case of decagon, $p(z)$ is a cubic polynomial.
We choose it as 
\[
p(z)=z^3-3\iLambda^2 z+u=(z-z_1)(z-z_2)(z-z_3).
\]
It is important to notice that we should essentially consider two cases
for the configurations of the roots $z_i$ ($i=1,2,3$) or for the
location of $u$ in the moduli space.
This is called the wall-crossing phenomenon in the literature. 

\subsubsection{Inside the wall of marginal stability}

\begin{figure}[tb]
  \begin{center}
    \includegraphics[width=15cm]{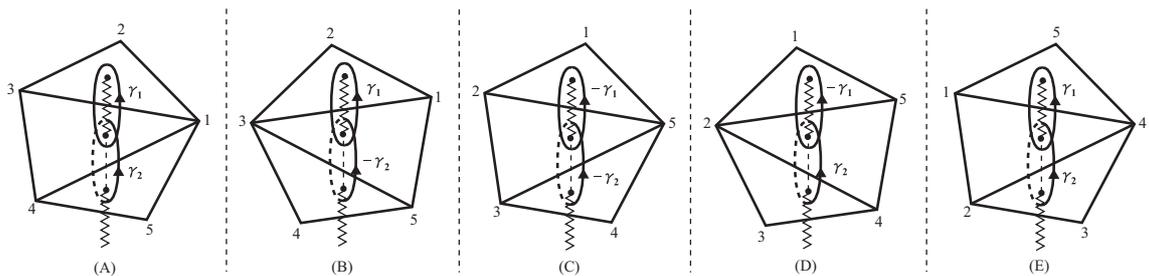}
  \end{center}
  \caption{Transition of WKB triangulations for the decagon when $u$ is
 inside the wall of marginal stability. We show the evolution of the WKB
 triangulations from $\arg \zeta=0$ to $\arg \zeta =2\pi$. There are
 four jumps in this case.}
  \label{fig:WKBlines}
\end{figure}

Let us first consider the case where $u$ is located inside the wall of
marginal stability (see Sec.~9.4.4 ``$N=3$'' in \cite{Gaiotto:2009hg}). 
Let $\gamma_1,\gamma_2$ denote cycles which encircle
the pair of branch points $[z_1,z_2]$, $[z_3,z_2]$,
respectively (see Figure~\ref{fig:WKBlines} (A)).
Given the phase of $\zeta$, one can draw the WKB lines
and determine the WKB triangulation. 
Figure~\ref{fig:WKBlines} schematically shows the evolution of the WKB
triangulations as the phase of $\zeta$ increases.
The WKB triangulations jumps discontinuously
when $\zeta$ crosses semi-infinite lines (so-called BPS rays).
We see that for generic $\zeta$ there always exist
two tetragons each of which respectively surround the 
edges $[z_1,z_2]$, $[z_3,z_2]$.
Once the WKB triangulation is given, we can define functions
$\chi_{\gamma_i}(\zeta)$ as the Fock--Goncharov coordinates for each
tetragon.
For example, let us consider the case where $\arg \zeta$ is in the
region (B) of Figure~\ref{fig:WKBlines}.
We define $\chi_{\gamma_i}$ ($i=1,2$) by
\begin{align}
\chi_{\gamma_1}&=
-\frac{(s_1\wedge s_2)(s_3\wedge s_5)}
     {(s_2\wedge s_3)(s_5\wedge s_1)}
=-\beta_4,\\
\chi_{\gamma_2}^{-1}&=
\chi_{-\gamma_2}=-\frac{(s_5\wedge s_1)(s_3\wedge s_4)}
     {(s_1\wedge s_3)(s_4\wedge s_5)}
=-\beta_2^{-1},
\end{align}
where we used the relation
$\chi_{-\gamma}(\zeta)=1/\chi_{\gamma}(\zeta)$.
Similarly one can define $\chi_{\gamma_i}$ ($i=1,2$) for other
regions. The definition of $\chi_{\gamma_i}$ ($i=1,2$) is summarized
in Table~\ref{tab:chi_dec_in}.

\begin{table}[tb]
 \caption{The definition of $\chi_{\gamma_1}$ and $\chi_{\gamma_2}$
 for the decagon when $u$ is inside the wall of marginal stability. The
 regions of $\arg \zeta$ correspond to those of
 Figure~\ref{fig:WKBlines}. \label{tab:chi_dec_in}}
 \begin{center}
  \begin{tabular}{cccccc}
    \hline
     $\arg \zeta$  &  (A)  &  (B)  &  (C)  &  (D)  &  (E)  \\
    \hline
     $\chi_{\gamma_1}$  &  $-\beta_5^{-1}$  &  $-\beta_4$ 
  &  $-\beta_4$  &  $-\beta_3^{-1}$  & $-\beta_3^{-1}$   \\
     $\chi_{\gamma_2}$  &  $-\beta_2$  &  $-\beta_2$  
  &  $-\beta_1^{-1}$  &  $-\beta_1^{-1}$  & $-\beta_5$   \\
    \hline
  \end{tabular}
 \end{center}
\end{table}

\begin{figure}[tb]
  \begin{center}
    \includegraphics[width=12cm]{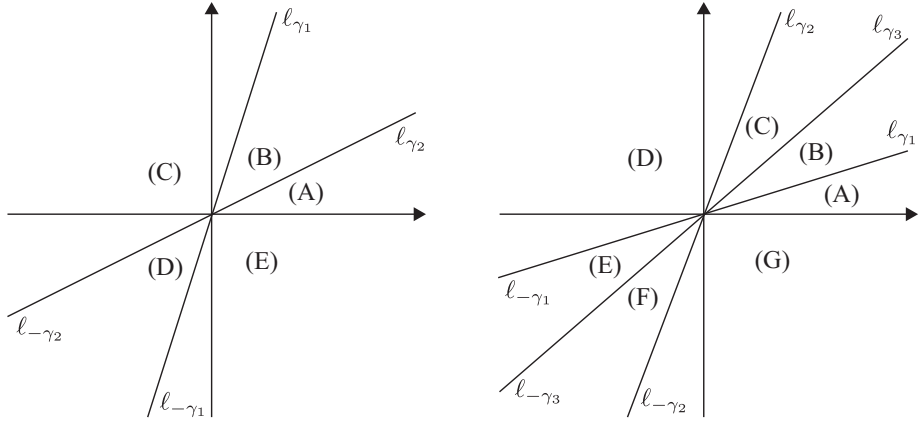}
  \end{center}
  \caption{
The BPS rays for the decagon solutions.
The left (right) figure shows the case where
$u$ is located inside (outside) the wall of marginal stability.}
  \label{fig:BPSrays.eps}
\end{figure}

Note that $\chi_{\gamma_i}$ defined in this way has the asymptotic
form \cite{Gaiotto:2009hg}
\eqb\label{chiasymp}
\chi_{\gamma_i}(\zeta)\simeq 
\exp\left(\frac{Z_i}{\zeta}+\bZ_i\zeta\right)
\eqe
for large $|\zeta|$ where
\eqb
Z_i=\oint_{\gamma_i}\sqrt{p(z)}dz.
\eqe
From the discontinuity data of $\chi_{\gamma_i}(\zeta)$
and the asymptotic form (\ref{chiasymp}),
we can immediately write down the integral equations for
$\chi_{\gamma_i}$,
\begin{align}
\log \chi_{\gamma_1}(\zeta) =\frac{Z_1}{\zeta}+\bar{Z}_1\zeta
&+\frac{1}{4\pi i} \int_{\ell_{\gamma_2}}
 \frac{d\zeta'}{\zeta'} \frac{\zeta'+\zeta}{\zeta'-\zeta}
 \log(1+\chi_{\gamma_2}(\zeta')) \notag \\
&-\frac{1}{4\pi i} \int_{\ell_{-\gamma_2}}
 \frac{d\zeta'}{\zeta'} \frac{\zeta'+\zeta}{\zeta'-\zeta}
 \log(1+\chi_{-\gamma_2}(\zeta')), \label{eq:Dec_int_in1} \\
\log \chi_{\gamma_2}(\zeta) =\frac{Z_2}{\zeta}+\bar{Z}_2\zeta 
&-\frac{1}{4\pi i} \int_{\ell_{\gamma_1}}
 \frac{d\zeta'}{\zeta'} \frac{\zeta'+\zeta}{\zeta'-\zeta}
 \log(1+\chi_{\gamma_1}(\zeta')) \notag \\
&+\frac{1}{4\pi i} \int_{\ell_{-\gamma_1}}
\frac{d\zeta'}{\zeta'} \frac{\zeta'+\zeta}{\zeta'-\zeta}
 \log(1+\chi_{-\gamma_1}(\zeta')), \label{eq:Dec_int_in2}
\end{align}
where the contours $\ell_{\gamma'}$ is chosen as
(see Figure~\ref{fig:BPSrays.eps})
\[
\ell_{\gamma'}: \frac{Z_{\gamma'}}{\zeta'} \in \mathbb{R}_- .
\]
%
Following Appendix E in \cite{Gaiotto:2008cd}, 
we can rewrite \eqref{eq:Dec_int_in1} and \eqref{eq:Dec_int_in2}
as the following TBA equations,
\eqb
\epsilon_1(\theta)\Eqn{=}2|Z_1|\cosh\theta
 -\int_{-\infty}^\infty \frac{d\theta'}{2\pi}
 \frac{1}{\cosh(\theta-\theta'-i\talpha)}
 \log(1+e^{-\epsilon_2(\theta')}), \label{eq:TBA5-1}\\
\epsilon_2(\theta)\Eqn{=}2|Z_2|\cosh\theta
 -\int_{-\infty}^\infty \frac{d\theta'}{2\pi}
 \frac{1}{\cosh(\theta-\theta'+i\talpha)}
 \log(1+e^{-\epsilon_1(\theta')}), \label{eq:TBA5-2}
\eqe
where we have introduced $\theta$ and
$\epsilon_k(\theta)\equiv\epsilon_{\gamma_k}(\theta)$ as
$Z_{k}=|Z_{k}|e^{i\alpha_{k}},\;
 \zeta=-e^{\theta+i \alpha_{k}},$
$\chi_{\gamma_k}(\zeta=-e^{\theta+i \alpha_{k}})
 =e^{-\epsilon_{\gamma_k}(\theta)}$,
and $\talpha \equiv \pi/2-(\alpha_1-\alpha_2)$.
We used the relations
$\epsilon_{-\gamma_k}(\theta)=\epsilon_{\gamma_k}(\theta)$.
These relations hold from the $\bbZ_2$-symmetry (\ref{Z2symmetry}).

If $u=0$, $Z_1$ and $Z_2$ are simply related to $\iLambda$,
\eqb
Z_1\Eqn{=}-2 \int_{-\sqrt{3}\iLambda}^0 dz \sqrt{p(z)}
=-\frac{\sqrt{\pi}\Gamma(\frac{3}{4})}{2\Gamma(\frac{9}{4})}
 (\sqrt{3} \iLambda)^{5/2}, \\
Z_2\Eqn{=}-2 \int_0^{\sqrt{3}\iLambda} dz \sqrt{p(z)} =iZ_1.
\eqe
Since $\alpha_1-\alpha_2=\pi/2$,
\eqref{eq:TBA5-1} and \eqref{eq:TBA5-2} yield
\eqb
\epsilon_1(\theta)\Eqn{=}2|Z|\cosh\theta
-\int_{-\infty}^\infty \frac{d\theta'}{2\pi}
 \frac{1}{\cosh(\theta-\theta')}
 \log(1+e^{-\epsilon_2(\theta')}), \label{eq:TBA5-3} \\
\epsilon_2(\theta)\Eqn{=}2|Z|\cosh\theta
-\int_{-\infty}^\infty \frac{d\theta'}{2\pi}
 \frac{1}{\cosh(\theta-\theta')}
 \log(1+e^{-\epsilon_1(\theta')}). \label{eq:TBA5-4}
\eqe
%

\subsubsection{Outside the wall of marginal stability}
\begin{figure}[tb]
  \begin{center}
    \includegraphics[width=15cm]{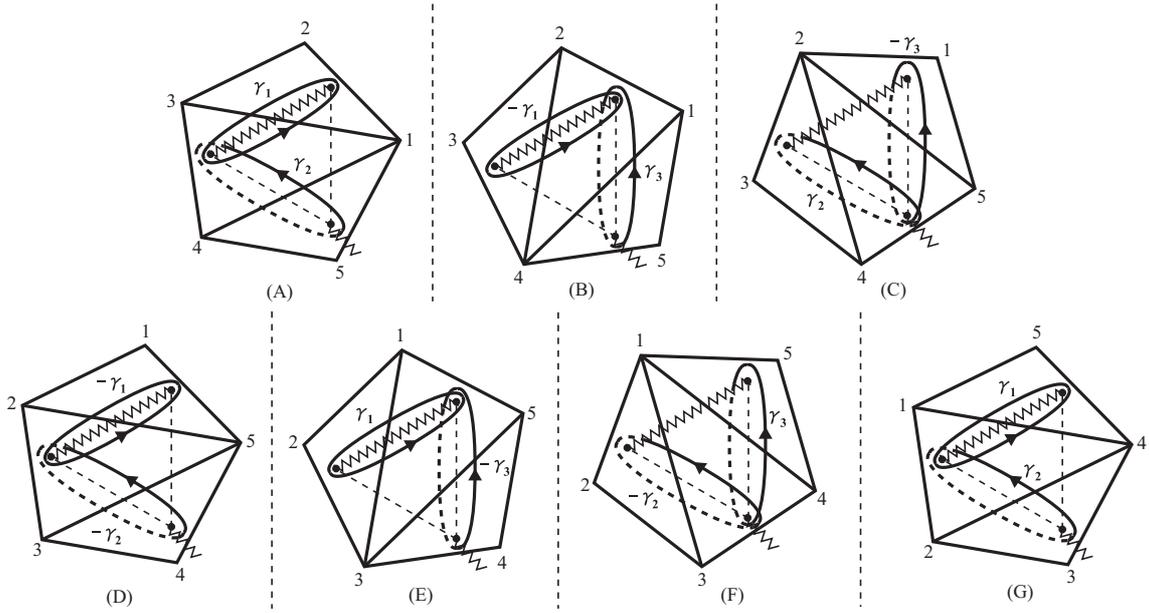}
  \end{center}
  \caption{Transition of WKB triangulations for the decagon when $u$
 is outside the wall of marginal stability. There are six jumps as
 $\arg \zeta$ varies from $0$ to $2\pi$.}
  \label{fig:wkb0.eps}
\end{figure}

Next let us consider the case where $u$ is located outside the wall of
marginal stability. In this case, as $\arg \zeta$ increases, the WKB
triangulations change as shown in Figure~\ref{fig:wkb0.eps}.
In the same way as the previous case, we define three functions
$\chi_{\gamma_i}(\zeta)$ $(i=1,2,3)$ from the WKB triangulations.
These are summarized in Table~\ref{tab:chi_dec_out}.
%
%
\begin{table}[tb]
 \caption{The definition of $\chi_{\gamma_i}(\zeta)$ $(i=1,2,3)$
 for the decagon when $u$ is outside the wall of marginal stability.
 \label{tab:chi_dec_out}}
 \begin{center}
  \begin{tabular}{cccccccc}
    \hline
   $\arg \zeta$ & (A) & (B) & (C) & (D) & (E) & (F) & (G) \\
    \hline
     $\chi_{\gamma_1}$  &  $-\beta_5^{-1}$ & $-\beta_5^{-1}$
  &  $\beta_1/\beta_3$  &  $-\beta_4$  &  $-\beta_4$
  &  $\beta_2/\beta_5$  & $-\beta_3^{-1}$   \\
     $\chi_{\gamma_2}$  &  $-\beta_2$  & $\beta_5/\beta_3$
  &  $-\beta_1^{-1}$  &  $-\beta_1^{-1}$  & $\beta_2/\beta_4$
  &  $-\beta_5$  & $-\beta_5$   \\
     $\chi_{\gamma_3}$ & $\beta_2/\beta_5$ & $-\beta_3^{-1}$
  &  $-\beta_3^{-1}$ & $\beta_4/\beta_1$ & $-\beta_2$
  &  $-\beta_2$ & $\beta_5/\beta_3$ \\
    \hline
  \end{tabular}
 \end{center}
\end{table}
Note that for all regions these functions satisfy the relation
$\chi_{\gamma_3}(\zeta)=
 \chi_{\gamma_1}(\zeta)\chi_{\gamma_2}(\zeta)$. 
We can write down the integral equations for
$\chi_{\gamma_i}(\zeta)$ $(i=1,2,3)$, and rewrite them as
the following TBA equations, 
\eqb
\epsilon_1(\theta)=2|Z_1|\cosh\theta
\Eqn{-}\int_{-\infty}^\infty \frac{d\theta'}{2\pi} 
 \frac{1}{\cosh(\theta-\theta'-i\talpha_{12})}
 \log(1+e^{-\epsilon_2(\theta')}) \notag \\
\Eqn{-}\int_{-\infty}^\infty \frac{d\theta'}{2\pi}
 \frac{1}{\cosh(\theta-\theta'-i\talpha_{13})}
 \log(1+e^{-\epsilon_3(\theta')}), \label{eq:TBA5-5}\\
\epsilon_2(\theta)=2|Z_2|\cosh\theta
\Eqn{-}\int_{-\infty}^\infty \frac{d\theta'}{2\pi}
 \frac{1}{\cosh(\theta-\theta'+i\talpha_{12})}
 \log(1+e^{-\epsilon_1(\theta')}) \notag \\
\Eqn{-}\int_{-\infty}^\infty \frac{d\theta'}{2\pi}
 \frac{1}{\cosh(\theta-\theta'+i\talpha_{32})}
 \log(1+e^{-\epsilon_3(\theta')}), \label{eq:TBA5-6}\\
\epsilon_3(\theta)=2|Z_3|\cosh\theta
\Eqn{-}\int_{-\infty}^\infty \frac{d\theta'}{2\pi}
 \frac{1}{\cosh(\theta-\theta'+i\talpha_{13})}
 \log(1+e^{-\epsilon_1(\theta')}) \notag \\
\Eqn{-}\int_{-\infty}^\infty \frac{d\theta'}{2\pi}
 \frac{1}{\cosh(\theta-\theta'-i\talpha_{32})}
 \log(1+e^{-\epsilon_2(\theta')}), \label{eq:TBA5-7}
\eqe
where $Z_3=Z_1+Z_2$ and
$\talpha_{ab}\equiv \pi/2-(\alpha_a-\alpha_b)$.

\subsection{Dodecagon solutions {\rm ($n=6$)}}

\begin{figure}[tb]
  \begin{center}
    \includegraphics[width=15cm]{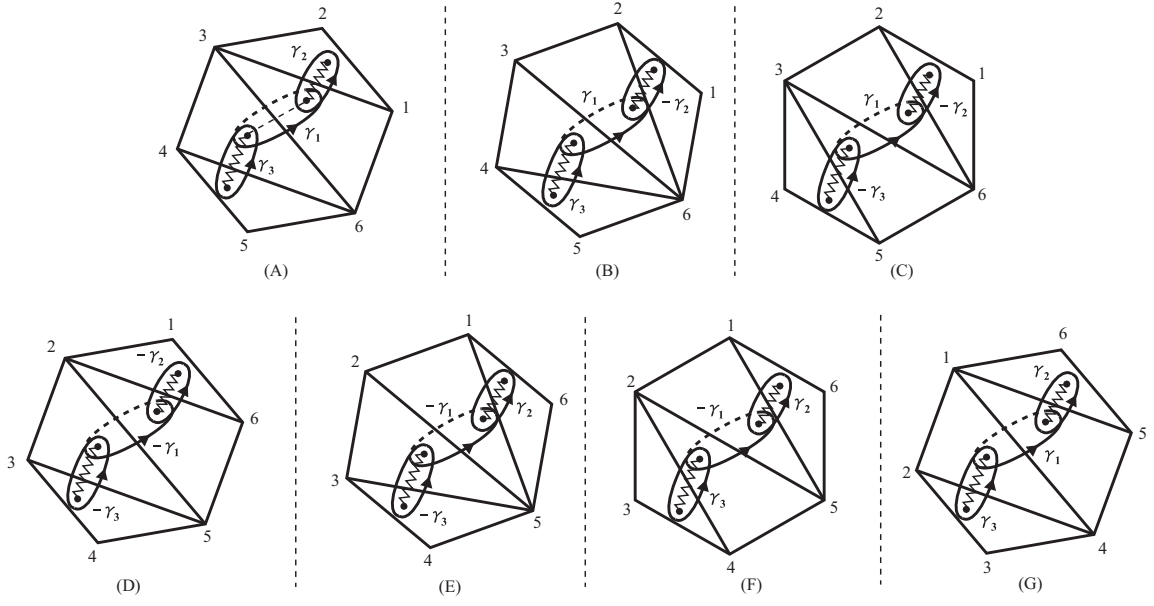}
  \end{center}
  \caption{Transition of WKB lines for the dodecagonal case.}
  \label{fig:wkb6.eps}
\end{figure}

In the dodecagonal case $n=6$,
the degree of the polynomial $p(z)$ is four, and we choose it as
\[
p(z)=z^4+4\iLambda^2 z^2+2mz+u
=(z-z_1)(z-z_2)(z-z_3)(z-z_4).
\]
From \eqref{BMrel}, the relations among $b$'s are given by
\eqb
&b_{j+2}b_{j+3}=1+\mu_j-\mu_jb_{j+5}b_{j+6},
\label{brel_dodec1}\\
&b_{j+1}+b_{j+3}-b_{j+1}b_{j+2}b_{j+3}= \mu_{j+1}b_{j+5}.
\label{brel_dodec2}
\eqe
As mentioned above, the wall-crossing phenomenon also occurs in this
case. Here we focus on the simplest case, {\it i.e.}, we only consider
the region where the number of BPS rays is the smallest
(three$+$three). The WKB triangulations evolve as in
Figure~\ref{fig:wkb6.eps}.
The definition of $\chi_{\gamma_i}$ ($i=1,2,3$)
is summarized in Table~\ref{tab:chi_dodec}.
\begin{table}[tb]
 \caption{The definition of $\chi_{\gamma_i}(\zeta)$ $(i=1,2,3)$
 for the dodecagonal case.\label{tab:chi_dodec}}
 \begin{center}
  \begin{tabular}{ccccc}
    \hline
     $\arg \zeta$  &  (A)  &  (B)  &  (C)  &  (D)   \\
    \hline
     $\chi_{\gamma_1}$  &  $-1/\mu b_2b_5$ & $-b_1/\mu b_5$
  &  $-b_1b_4/\mu$  &  $-b_1b_4/\mu$     \\
     $\chi_{\gamma_2}$  &  $\mu(1-b_4b_5)$  & $\mu(1-b_4b_5)$
  &  $\mu(1-b_4b_5)$  &  $\mu/(1-b_3b_4)$     \\
     $\chi_{\gamma_3}$ & $-(1-b_4b_5)$  & $-(1-b_4b_5)$
  &  $-(1-b_4b_5)$  &  $-1/(1-b_3b_4)$   \\
    \hline
    \hline
    $\arg \zeta$  &  (E)  &  (F)  &  (G)  &    \\
    \hline
    $\chi_{\gamma_1}$  &  $\mu b_4/b_6$ & $-\mu/b_3b_6$ 
  & $-\mu/b_3b_6$ & \\
    $\chi_{\gamma_2}$  &  $\mu/(1-b_3b_4)$ & $\mu/(1-b_3b_4)$
  & $\mu(1-b_2b_3)$ &     \\
    $\chi_{\gamma_3}$ & $-1/(1-b_3b_4)$ & $-1/(1-b_3b_4)$ 
  & $-(1-b_2b_3)$ &    \\
	\hline
  \end{tabular}
 \end{center}
\end{table}
The discontinuity of
$\chi_{\gamma_i}$ can be expressed by the remaining two functions
using \eqref{brel_dodec1} and \eqref{brel_dodec2}. 
When $\arg \zeta$ crosses the BPS ray from (C) to (D) in Table
\ref{tab:chi_dodec}, for example, the ratio of the discontinuity of
$\chi_{\gamma_2}$ is evaluated as
\begin{align}
\frac{\chi_{\gamma_2}^+}{\chi_{\gamma_2}^-}
=\frac{1}{(1-b_3b_4)(1-b_4 b_5)}
=\frac{1}{1-\mu^{-1}b_1b_4}=(1+\chi_{\gamma_1})^{-1}.
\end{align}
Thus we finally obtain the TBA equations,
\eqb
\epsilon_1(\theta)=2|Z_1|\cosh\theta
\Eqn{-}\int_{-\infty}^\infty \frac{d\theta'}{2\pi}
 \frac{1}{\cosh(\theta-\theta'-i\talpha_{12})}
 \log(1+e^{-\epsilon_2(\theta')}) \notag \\
\Eqn{-}\int_{-\infty}^\infty \frac{d\theta'}{2\pi}
 \frac{1}{\cosh(\theta-\theta'-i\talpha_{13})}
 \log(1+e^{-\epsilon_3(\theta')}), \label{eq:TBA6-1}\\
\epsilon_2(\theta)=2|Z_2|\cosh\theta
\Eqn{-}\int_{-\infty}^\infty \frac{d\theta'}{2\pi}
 \frac{1}{\cosh(\theta-\theta'+i\talpha_{12})}
 \log(1+e^{-\epsilon_1(\theta')}), \label{eq:TBA6-2}\\
\epsilon_3(\theta)=2|Z_3|\cosh\theta
\Eqn{-}\int_{-\infty}^\infty \frac{d\theta'}{2\pi}
 \frac{1}{\cosh(\theta-\theta'+i\talpha_{13})}
 \log(1+e^{-\epsilon_1(\theta')}). \label{eq:TBA6-3}
\eqe
If the moduli parameter $u$ crosses the wall of marginal stability, the
integral equations should be modified as well as the decagonal
case. Although we can derive the TBA equations for such cases from the
WKB data in the same way, we do not write them down explicitly here.

\subsection{Integral equations for general $2n$-gon}

Until now, we have focused on the two special polygons: decagon and
dodecagon. The integral equations derived there 
have the same forms as those
in \cite{Gaiotto:2008cd}. Thus it is natural to
expect that these are true for general $2n$-gons. Here we rewrite the
integral equations in \cite{Gaiotto:2008cd} for our interested
situations. We will identify them with the TBA equations for the
homogeneous sine-Gordon models associated with the coset CFTs later.

Our starting equations are the followings,
\[
\log \chi_{\gamma_k}(\zeta)
=\frac{Z_{\gamma_k}}{\zeta}+\bar{Z}_{\gamma_k}\zeta
-\frac{1}{4\pi i}\sum_{\gamma'} \sf{\gamma_k}{\gamma'}
\int_{\ell_{\gamma'}} \frac{d\zeta'}{\zeta'}
\frac{\zeta'+\zeta}{\zeta'-\zeta}\log(1+\chi_{\gamma'}(\zeta')).
\label{eq:int_eq}
\]
For the $2n$-gon, $\gamma'$ in the sum runs over
$\pm\gamma_1,\pm\gamma_2,\ldots,\pm\gamma_{n-3}$.\footnote{
We focus on the simplest region of the moduli space.} 
The contour $\ell_{\gamma'}$ is chosen as
\[
\ell_{\gamma'}: \frac{Z_{\gamma'}}{\zeta'} \in \mathbb{R}_- \,.
\]
By combining the terms for $\gamma'=\gamma_k$ and
$\gamma'=-\gamma_k$ and using the $\bbZ_2$-symmetry,
which is inherent in the present $AdS_3$ system,
we obtain the following simple integral equations,
\[
\epsilon_k(\theta)=2|Z_k|\cosh\theta
-\sum_{l=1}^{n-3}\int_{-\infty}^\infty \frac{d\theta'}{2\pi}
 \frac{-i\sf{\gamma_k}{\gamma_l}}
      {\sinh(\theta-\theta'+i\alpha_k-i\alpha_l)}
 \log\left(1+e^{-\epsilon_l(\theta')}\right)\,,
\label{eq:TBA}
\]
where $Z_k \equiv Z_{\gamma_k}$.
Note that one can reproduce
(\ref{eq:TBA5-1})--(\ref{eq:TBA5-2}),
(\ref{eq:TBA6-1})--(\ref{eq:TBA6-3})
from (\ref{eq:TBA}).
The discussion outside the wall of marginal stability is similar.

\subsection{TBA equations of the homogeneous sine-Gordon model}

The homogeneous sine-Gordon models 
\cite{FernandezPousa:1996hi,FernandezPousa:1997zb} are 
a class of two-dimensional integrable models generalizing
the sine-Gordon model. They are obtained by integrable perturbations
of  conformal field theories\footnote{
Regarding these, see also \cite{IntPerturbation}.}
corresponding to $\grp{G}_k$-parafermions, or
cosets $\grp{G}_k/[\grp{U}(1)]^{r_\alg{g}}$ \cite{parafermions},
where $\grp{G}$ is a simple compact Lie group with
Lie algebra $\alg{g}$, and $r_\alg{g}$ is the rank of $\alg{g}$.
An integer $k$ is the level of affine Lie algebra $\hat{\alg{g}}$. 
The S-matrices describing the  models for  simply laced $\grp{G}$'s
are proposed in \cite{Miramontes:1999hx}.

For the minimal surfaces in $AdS_3$, it turns out that the case 
of the
$\grp{SU}(N)_2/[\grp{U}(1)]^{N-1}$ coset is relevant,
which is discussed in detail
in \cite{CastroAlvaredo:2000nr}.
An explicit form of the non-trivial part of the 
S-matrix  in this case is given by
\eqb
S_{ab}(\theta) = (-1)^{\delta_{ab}}\left[ c_a \tanh\frac{1}{2}
  \left(\theta+\sigma_{ab} -\frac{\pi}{2}i\right) \right]^{I_{ab}} .
\eqe 
Here, $a=1,\ldots, N-1$ labels the particles
corresponding to each simple root with
mass $m_a$, $I_{ab}$ is the incidence matrix, $c_a$ are constants
and $\sigma_{ab} = - \sigma_{ba}$ are some parameters.

Following the standard procedure, one finds the TBA equations 
from this S-matrix with inverse temperature $R$:
\eqb\label{TBAHSG}
  \epsilon_a(\theta) = m_a R \cosh \theta -
  \sum_b \int \frac{d\theta'}{2\pi}
  \frac{i I_{ab}}{\sinh(\theta-\theta'+\sigma_{ab}+\frac{\pi}{2}i)}
  \log(1+e^{-\epsilon_b}).
\eqe
Now it is clear that the TBA equations from
the $\grp{SU}(N)_2/[\grp{U}(1)]^{N-1}$ homogeneous 
sine-Gordon model coincide with those in (\ref{eq:TBA})
under the identifications $n-2\leftrightarrow N$, 
$2|Z_a|\leftrightarrow m_a R $, $ \langle \gamma_a,
 \gamma_b\rangle \leftrightarrow \epsilon_{ab}I_{ab} $, and
$i(\alpha_a-\alpha_b)\leftrightarrow\sigma_{ab}+\frac{\pi}{2}i$.
Here $\epsilon_{ab}=-\epsilon_{ba}=\pm 1$.

The TBA equations for the general homogeneous sine-Gordon
models can be derived from the S-matrices
in \cite{Miramontes:1999hx}. 
It would be  of interest to see the relevance 
to the minimal surfaces in $AdS_5$ and $AdS_4$. 
We comment on this point in the next section.

\section{Regularized area and free energy}

From the solutions to the TBA equations in sect.~3.1,~3.2, 
one can extract the physical quantities following
\cite{Alday:2009yn,Alday:2009dv}.
As we argue shortly, we expect that this is the case for
the TBA equations (\ref{eq:TBA}) or (\ref{TBAHSG}) with general $n$.
First, the cross-ratios of the $AdS_3$ boundary coordinates 
$x^\pm_{ij}x^\pm_{kl}/x^\pm_{ik}x^\pm_{jl}$
are given by
$(s_i\wedge s_j)(s_k\wedge s_l)/(s_i\wedge s_k)(s_j\wedge s_l)$
evaluated at $\zeta = 1$ and $\zeta=i$, respectively.
They are in turn read off from $\chi$'s and $\beta$'s. 
Second,
the area of the minimal surface, representing the scattering
amplitude, is decomposed as
\eqb
  A = A_{\rm sinh} + 4 \int d^2w , \quad
  A_{\rm sinh} = 4 \int d^2z \left(e^{2\alpha}-\sqrt{p\bar{p}}\right), 
\eqe
where $\int d^2 w$ is divergent and should be regularized.
As for the finite piece $A_{\rm sinh}$,  the  Poisson brackets among 
the Fock--Goncharov coordinates \cite{Gaiotto:2008cd} 
imply that the relation \cite{Alday:2009yn} between $A_{\rm sinh}$
and the free energy of the
TBA system, if any, generically holds:
\eqb\label{AF}
   A_{\rm sinh} = F + c_n.
\eqe
The constant term  $c_n$ is fixed  by 
considering the limit where the zeros of $p(z)$ become far apart from
each other \cite{Alday:2009yn}: In this limit, the solution is regarded
as a superposition of single-zero solutions. Since each single-zero
solution corresponds to the hexagon solution,
one has $A_{\rm sinh} \to (n-2)A_{\rm sinh}(n=3)$
with $A_{\rm sinh}(n=3) = 7\pi/12$. Therefore,
\eqb
  c_n = \frac{7}{12} (n-2) \pi.
\eqe

\subsection{CFT limit and coset models}
It is often the case that massive integrable models are obtained  by 
perturbing conformal field theories. 
The identification of the conformal model is useful to analyze
the TBA system. 

In the previous section, we found that the TBA equations for the
decagonal and the dodecagonal solutions are identified with  those of
the homogeneous  sine-Gordon model which are obtained by perturbing the
conformal field theory associated with the coset \cite{Bagger:1988}
\eqb\label{coset}
  \frac{\grp{SU}(n-2)_2}{[\grp{U}(1)]^{n-3}}
  \simeq \frac{[\grp{SU}(2)_1]^{n-2}}{\grp{SU}(2)_{n-2}},
\eqe
with $n=5,6$.
From the further identification with the TBA-like equations in
(\ref{eq:TBA}),
we expect that the general $2n$-gon solutions in $AdS_3$ are described
by the TBA equations of the above homogeneous sine-Gordon model
with general $n$.

In fact, as obvious from the right-hand side of (\ref{coset}), the
number of the degrees of freedom in this systems is
$(n-2)-1 = (2n-6)/2$, which matches 
the number of the independent cross ratios. We note here that 
the left and the right sectors are described by 
the same Hitchin system in the $AdS_3$ case.

Moreover, one finds a precise agreement in the conformal limit
between $A_{\rm sinh}$ from the minimal surfaces
and the free energy $F$
from the TBA equations of the homogeneous sine-Gordon model.
On the minimal surface  side, the solution in the conformal limit
reduces to the regular polygon
solution, where $A_{\rm sinh}$ is evaluated as \cite{Alday:2009yn}
 \eqb\label{Asinh}
   A_{\rm sinh} = \frac{\pi}{4n} (3n^2-8n +4).
 \eqe
On the TBA side, the free energy in the conformal limit is obtained by 
setting $m_a =0$ and is given by
the ground state energy of the corresponding
conformal model\cite{Zamolodchikov:1989cf}.
Since the coset model 
\eqb\label{coset2}
  \frac{\grp{SU}(K)_k}{[\grp{U}(1)]^{K-1}}
  \simeq \frac{[\grp{SU}(k)_1]^{K}}{\grp{SU}(k)_{K}}
\eqe
has the central charge $c = (k-1)K(K-1)/(k+K)$,
the free energy in our case is
\eqb\label{Fc}
  F = \frac{\pi}{6} c = \frac{\pi}{6n}(n-2)(n-3).
\eqe
Taking into account the constant term in (\ref{AF}), 
we find that 
\eqb
  F + c_n  = \frac{\pi}{4n} (3n^2-8n +4),
\eqe
which is in precise agreement  with (\ref{Asinh}).
One can also derive (\ref{Fc}) directly by starting from the
TBA equations of the homogeneous sine-Gordon model in (\ref{TBAHSG}) 
\cite{Bazhanov:1989yk,Kirillov:1987}.\footnote{
For $n=5$ one can also check that by starting from
(\ref{eq:TBA5-5})--(\ref{eq:TBA5-7}).}

Finally, we would like to comment on the case of $AdS_5$. 
In \cite{Alday:2009dv},  it was shown that the hexagon solution
in $AdS_5$ is described by the $A_3$ TBA system, which corresponds
to $k=4, K=2$ in (\ref{coset2}).
From a consideration on the degrees of freedom 
and the symmetry of the Hitchin system, we expect that the 
$m$-gon solution in $AdS_5$ is described by the TBA equations
of the homogeneous sine-Gordon model  
corresponding to the coset 
\eqb
\frac{\grp{SU}(m-4)_4}{[\grp{U}(1)]^{m-5}}
\simeq \frac{[\grp{SU}(4)_1]^{m-4}}{\grp{SU}(4)_{m-4}},
\eqe
in a region of marginal stability.
The coset model has the central charge $3(m-4)(m-5)/m$.
This should be reproduced from the regular polygon solutions.

\section{Conclusions}

In this paper we studied the classical open string solutions with
a null polygonal boundary in $AdS_3$.
We derived in full detail the set of integral equations for the
decagonal and the dodecagonal solutions. 
These integral equations were identified with the TBA equations of the 
homogeneous sine-Gordon model, whose CFT limits are the generalized
parafermion models. We also observed general correspondence between
the polygonal solutions in $AdS_n$ and generalized parafermions.

Since the deformations around the CFT point of homogeneous sine-Gordon
models have rich structure
\cite{CastroAlvaredo:2000nr,CastroAlvaredo:1999em,Dorey:2004qc},
it is interesting to study the remainder functions around
the CFT point. It is also interesting to study supersymmetric
extensions and quantum corrections.

\paragraph{Note Added:}
During the preparation of this paper, we have noticed the paper 
by Alday, Maldacena, Sever and Vieira\cite{Alday:2010vh},
which considerably overlaps with the present work.
In particular, they also present the TBA equations for
general $2n$-gons in the $AdS_3$ case at the end of sect.~3.5,
which coincide with ours (\ref{eq:TBA})
under appropriate identification of functional variables.
According to their results, the regular $m$-gon solutions in $AdS_5$
with $\mu=1$ is found to be consistent with the central charge of the
generalized parafermions.
The structure of the Y-system is also in accord with
the spectrum of the homogeneous sine-Gordon model.

\vspace{3ex}

\begin{center}
  {\bf Acknowledgments}
\end{center}

We would like to thank J.~Suzuki for useful discussions.
The work of K.~I., K.~S. and Y.~S. is supported in part by Grant-in-Aid
for Scientific Research from the Japan Ministry of Education, Culture, 
Sports, Science and Technology.
The work of Y.~H. is supported by JSPS research fellowships
for young scientists.


%
%
\def\thebibliography#1{\list
 {[\arabic{enumi}]}{\settowidth\labelwidth{[#1]}\leftmargin\labelwidth
  \advance\leftmargin\labelsep
  \usecounter{enumi}}
  \def\newblock{\hskip .11em plus .33em minus .07em}
  \sloppy\clubpenalty4000\widowpenalty4000
  \sfcode`\.=1000\relax}
 \let\endthebibliography=\endlist
\vspace{3ex}
\begin{center}
 {\bf References}
\end{center}
\par \vspace*{-2ex}


\begin{thebibliography}{999}
\parskip=-2.5pt
%

\bibitem{Dr}
J.M.~Drummond, G.P.~Korchemsky and E. Sokatchev,
Nucl.\ Phys.\ B {\bf 795} (2008) 385
[arXiv:0707.0243[hep-th]].\\
A.~Brundhuber, P.~Heslop and G.~Travaglini,
Nucl.\ Phys.\ B {\bf 794} (2008) 231
[arXiv:0707:1153[hep-th]].\\
J.M.~Drummond, J.~Henn, G.P.~Korchemsky and E.~Sokatchev,
Nucl.\ Phys.\ B {\bf 795} (2008) 52
[arXiv.0709.2368[hep-th]];
Nucl.\ Phys.\ B {\bf 826} (2010) 337
[arXiv:0712.1223[hep-th]];
Nucl.\ Phys.\ B {\bf 828} (2010) 317
[arXiv:0807.1095 [hep-th]].
\\
R.~Ricci, A.~A.~Tseytlin and M.~Wolf,
  JHEP {\bf 0712} (2007) 082
  [arXiv:0711.0707 [hep-th]].
\\
 N.~Berkovits and J.~Maldacena,
  JHEP {\bf 0809} (2008) 062
  [arXiv:0807.3196 [hep-th]].\\
N.~Beisert, R.~Ricci, A.~A.~Tseytlin and M.~Wolf,
  Phys.\ Rev.\  D {\bf 78} (2008) 126004
  [arXiv:0807.3228 [hep-th]].

\bibitem{Alday:2007hr}
  L.~F.~Alday and J.~M.~Maldacena,
  JHEP {\bf 0706} (2007)  064
  [arXiv:0705.0303 [hep-th]].

\bibitem{various}
S.~Abel, S.~Forste and V.V.~Khoze,
JHEP {\bf 0802} (2008) 042 [arXiv:[0705.2113]].\\
E.~I. Buchbinder,
Phys.\ Lett.\ B {\bf 654} (2007) 46 [arXiv: 0706.2015[hep-th]].
\\
S.~Ryang,
Phys.\ Lett.\ B {\bf 659} (2008) 894 [arXiv:0710.1673[hep-th]];
  arXiv:0910.4796 [hep-th].
\\
A.~Popolitov,
arXiv:0710.2073[hep-th].\\
G.~Yang,
JHEP {\bf 0803} (2008) 010
[arXiv: 0711.2828[hep-th]].\\
A.~Jevicki, K.~Jin, C.~Kalousios and A.~Volovich,
JHEP {\bf 0803} (2008) 032 [arXiv:0712.1193[hep-th]].\\
Z.~Komargodski,
JHEP {\bf 0805} (2008) 019
[arXiv:0801.3274[hep-th]].\\
  R.~C.~Brower, H.~Nastase, H.~J.~Schnitzer and C.~I.~Tan,
  Nucl.\ Phys.\  B {\bf 814} (2009) 293
  [arXiv:0801.3891 [hep-th]].
A.~Mironov, A.~Morozov, and T.~N.~Tomaras,
  JHEP {\bf 0711} (2007) 021 [arXiv:0708.1625 [hep-th]];
Phys. Lett. {\bf B659} (2008) 723 [arXiv:0711.0192[hep-th]]\\
 H.~Itoyama, A.~Mironov and A.~Morozov,
  Nucl.\ Phys.\  B {\bf 808} (2009) 365
  [arXiv:0712.0159 [hep-th]].
\\
 H.~Itoyama and A.~Morozov,
  Prog.\ Theor.\ Phys.\  {\bf 120} (2008) 231
  [arXiv:0712.2316 [hep-th]].
\\
D.~Galakhov, H.~Itoyama, A.~Mironov and A.~Morozov,
  Nucl.\ Phys.\  B {\bf 823} (2009) 289
  [arXiv:0812.4702 [hep-th]].
\\
Z.~Komargodski and S.~S.~Razamat,
JHEP {\bf 0801} (2008) 044
  [arXiv:0707.4367 [hep-th]].\\
J.~McGreevy and A.~Sever,
JHEP {\bf 0802} (2008) 015 [arXiv: 0710.0393[hep-th]];
  JHEP {\bf 0808} (2008) 078
  [arXiv:0806.0668 [hep-th]].
\\
 D.~Astefanesei, S.~Dobashi, K.~Ito and H.~Nastase,
JHEP {\bf 0710} (2007) 077 [arXiv:0710.1684[hep-th]].
\\
K.~Ito, H.~Nastase and K.~Iwasaki,
  Prog.\ Theor.\ Phys.\  {\bf 120} (2008) 99
  [arXiv:0711.3532 [hep-th]].
\\
Y.~Oz, S.~Theisen and S.~Yankielowicz,
Phys.\ Lett.\ B {\bf 662} (2008) 297 [arXiv:0712.3491 [hep-th]].
\\
M.~Kruczenski,
JHEP {\bf 0212} (2002) 024
[hep-th/0210115].
\\
S.~Dobashi, K.~Ito and K.~Iwasaki,
JHEP {\bf 0807} (2008) 088
[arXiv:0805.3594 [hep-th]].
\\
S.~Dobashi and K.~Ito,
  Nucl.\ Phys.\  B {\bf 819} (2009) 18
  [arXiv:0901.3046 [hep-th]].
\\
  A.~Jevicki and K.~Jin,
  JHEP {\bf 1003} (2010) 028
  [arXiv:0911.1107 [hep-th]].
\\
 K.~Sakai and Y.~Satoh,
  JHEP {\bf 0910} (2009) 001
  [arXiv:0907.5259 [hep-th]].
\\
H.~Dorn, G.~Jorjadze and S.~Wuttke,
  JHEP {\bf 0905} (2009) 048
  [arXiv:0903.0977 [hep-th]].
\\
H.~Dorn,
  JHEP {\bf 1002} (2010) 013
  [arXiv:0910.0934 [hep-th]].
\\
H.~Dorn, N.~Drukker, G.~Jorjadze and C.~Kalousios,
  arXiv:0912.3829 [hep-th].

\bibitem{Sakai:2010eh}
  K.~Sakai and Y.~Satoh,
  JHEP {\bf 1003} (2010) 077
  [arXiv:1001.1553 [hep-th]].

\bibitem{BDS}
Z.~Bern, L.~J.~Dixon and V.~A.~Smirnov,
Phys.\ Rev.\  D {\bf 72} (2005) 085001
[arXiv:hep-th/0505205].

\bibitem{Dr2}
J.M.~Drummond, J.~Henn, G.P.~Korchemsky and E.~Sokatchev,
Nucl.\ Phys.\ B {\bf 815} (2009) 142
        [arXiv:0803.1466[hep-th]].\\
Z.~Bern, L.J.~Dixon, D.A.~Kosower, R.~Roiban, M.~Spradlin, C.~Vergu
and A.~Volovich,
Phys.\ Rev.\ D {\bf 78} (2008) 045007
[arXiv:0803.1465[hep-th]].

\bibitem{nume}
C.~Anastasiou, A.~Brandhuber, P.~Heslop, V.~V.~Khoze, B.~Spence
and G.~Travaglini,
  JHEP {\bf 0905} (2009) 115
  [arXiv:0902.2245 [hep-th]].
\\
A.~Brandhuber, P.~Heslop, V.~V.~Khoze and G.~Travaglini,
  JHEP {\bf 1001}  (2010) 050
  [arXiv:0910.4898 [hep-th]].

\bibitem{Smir}
  V.~Del Duca, C.~Duhr and V.~A.~Smirnov,
  JHEP {\bf 1003} (2010) 099
  [arXiv:0911.5332 [hep-ph]].

\bibitem{Alday:2009yn}
  L.~F.~Alday and J.~Maldacena,
  JHEP {\bf 0911} (2009) 082
  [arXiv:0904.0663 [hep-th]].

\bibitem{Gaiotto:2008cd}
  D.~Gaiotto, G.~W.~Moore and A.~Neitzke,
  arXiv:0807.4723 [hep-th].

\bibitem{Gaiotto:2009hg}
  D.~Gaiotto, G.~W.~Moore and A.~Neitzke,
  arXiv:0907.3987 [hep-th].

\bibitem{Alday:2009dv}
  L.~F.~Alday, D.~Gaiotto and J.~Maldacena,
  arXiv:0911.4708 [hep-th].

\bibitem{Burrington:2009bh}
  B.~A.~Burrington and P.~Gao,
  arXiv:0911.4551 [hep-th].

\bibitem{TBAinAdSCFT}
  N.~Gromov, V.~Kazakov and P.~Vieira,
  Phys.\ Rev.\ Lett.\  {\bf 103} (2009) 131601
  [arXiv:0901.3753 [hep-th]].
\\
  D.~Bombardelli, D.~Fioravanti and R.~Tateo,
  J.\ Phys.\ A {\bf 42} (2009) 375401
  [arXiv:0902.3930 [hep-th]].
\\
  N.~Gromov, V.~Kazakov, A.~Kozak and P.~Vieira,
  Lett.\ Math.\ Phys.\  {\bf 91} (2010) 265
  [arXiv:0902.4458 [hep-th]].
\\
  G.~Arutyunov and S.~Frolov,
  JHEP {\bf 0905} (2009) 068
  [arXiv:0903.0141 [hep-th]].

\bibitem{Zamolodchikov:1989cf}
  Al.~B.~Zamolodchikov,
  Nucl.\ Phys.\  B {\bf 342} (1990) 695.

\bibitem{FernandezPousa:1996hi}
  C.~R.~Fernandez-Pousa, M.~V.~Gallas, T.~J.~Hollowood
  and J.~L.~Miramontes,
  Nucl.\ Phys.\  B {\bf 484} (1997) 609
  [arXiv:hep-th/9606032].

\bibitem{FernandezPousa:1997zb}
  C.~R.~Fernandez-Pousa, M.~V.~Gallas, T.~J.~Hollowood
  and J.~L.~Miramontes,
  Nucl.\ Phys.\  B {\bf 499} (1997) 673
  [arXiv:hep-th/9701109].

\bibitem{Fock:2006}
  V.~Fock and A.~Goncharov,
  Publ.\ Math.\ Inst.\ Hautes\ \'Etudes\ Sci.\ {\bf 103} (2006) 1--211
  [arXiv:math/0311149].

\bibitem{DoreyTateo}
  P.~Dorey and R.~Tateo,
  J.\ Phys.\ A  {\bf 32} (1999) L419
  [arXiv:hep-th/9812211];
  Nucl.\ Phys.\  B {\bf 563} (1999) 573
  [Erratum-ibid.\  B {\bf 603} (2001) 581]
  [arXiv:hep-th/9906219].

\bibitem{Gliozzi:1995wq}
  F.~Gliozzi and R.~Tateo,
  Int.\ J.\ Mod.\ Phys.\  A {\bf 11} (1996) 4051
  [arXiv:hep-th/9505102].

\bibitem{TBAODE}
  V.~V.~Bazhanov, S.~L.~Lukyanov and A.~B.~Zamolodchikov,
  J.\ Statist.\ Phys.\  {\bf 102} (2001) 567
  [arXiv:hep-th/9812247].
\\
  P.~Dorey, C.~Dunning, D.~Masoero, J.~Suzuki and R.~Tateo,
  Nucl.\ Phys.\  B {\bf 772} (2007) 249
  [arXiv:hep-th/0612298].

%
\bibitem{IntPerturbation}
  I.~Bakas,
  Int.\ J.\ Mod.\ Phys.\  A {\bf 9} (1994) 3443
  [arXiv:hep-th/9310122].
\\
  Q.~H.~Park,
  Phys.\ Lett.\  B {\bf 328} (1994) 329
  [arXiv:hep-th/9402038].
\\
  I.~Bakas, Q.~H.~Park and H.~J.~Shin,
  Phys.\ Lett.\  B {\bf 372} (1996) 45
  [arXiv:hep-th/9512030].

%
\bibitem{parafermions}
  D.~Gepner,
  Nucl.\ Phys.\  B {\bf 290} (1987) 10.
\\
  D.~Gepner and Z.~Qiu,
  Nucl.\ Phys.\  B {\bf 285} (1987) 423.
\\
  V.~A.~Fateev and A.~B.~Zamolodchikov,
  Sov.\ Phys.\ JETP {\bf 62} (1985) 215
  [Zh.\ Eksp.\ Teor.\ Fiz.\  {\bf 89} (1985) 380].

\bibitem{Miramontes:1999hx}
  J.~L.~Miramontes and C.~R.~Fernandez-Pousa,
  Phys.\ Lett.\  B {\bf 472} (2000) 392
  [arXiv:hep-th/9910218].

\bibitem{CastroAlvaredo:2000nr}
  O.~A.~Castro-Alvaredo and A.~Fring,
  Phys.\ Rev.\  D {\bf 64} (2001) 085007
  [arXiv:hep-th/0010262].

\bibitem{Bagger:1988}
  J.~Bagger and D.~Nemeschansky,
  ``Coset construction of chiral algebras,''
  Proceedings of the Maryland Superstring Workshop,
  Eds.~G.~Gates et al.,
  World Scientific, Singapore (1988).

%
\bibitem{Bazhanov:1989yk}
  V.~Bazhanov and N.~Reshetikhin,
  J.\ Phys.\ A  {\bf 23} (1990) 1477.

\bibitem{Kirillov:1987}
  A.~N.~Kirillov,
Zap.\ Nauchn.\ Semin.\ Leningr.\ Otdel.\ Mat.\ Inst.\ {\bf 164}
 (1987) 121
[J.\ Sov.\ Math.\  {\bf 47} (1989) 2450].

\bibitem{CastroAlvaredo:1999em}
  O.~A.~Castro-Alvaredo, A.~Fring, C.~Korff and J.~L.~Miramontes,
  Nucl.\ Phys.\  B {\bf 575} (2000) 535
  [arXiv:hep-th/9912196].

\bibitem{Dorey:2004qc}
  P.~Dorey and J.~L.~Miramontes,
  Nucl.\ Phys.\  B {\bf 697} (2004) 405
  [arXiv:hep-th/0405275].

\bibitem{Alday:2010vh}
  L.~F.~Alday, J.~Maldacena, A.~Sever and P.~Vieira,
  arXiv:1002.2459 [hep-th].

%
\end{thebibliography}
\end{document}